\documentclass[aps,pxd,twocolumn,reprint,nofootinbib]{revtex4-2}
\usepackage{graphicx}  

\usepackage{dcolumn}  
\usepackage{bm}        
\usepackage{amssymb}
\usepackage{amsmath} 
\usepackage{epstopdf} 
\usepackage{amsfonts,float}
\usepackage{fancyhdr}
\usepackage{pifont}
\usepackage[hidelinks]{hyperref} 
\usepackage{bbm}
\usepackage{tensor}
\usepackage{mathrsfs}
\usepackage{xcolor}
\usepackage{physics}
\usepackage{fancyvrb} 
\usepackage{subcaption}
\usepackage{mathtools}

\usepackage{color} 

\usepackage[titles]{tocloft}
		\newcommand{\ddp}{\delta\text{-}\delta'}
		
		\newcommand{\bea}{\begin{eqnarray}} 
		\newcommand{\eea}{\end{eqnarray}}
		\newcommand{\beq}{\begin{equation}} 
		\newcommand{\eeq}{\end{equation}}

\newcommand{\qed}{\nobreak \ifvmode \relax \else
      \ifdim\lastskip<1.5em \hskip-\lastskip
      \hskip1.5em plus0em minus0.5em \fi \nobreak
      \vrule height0.75em width0.5em depth0.25em\fi}

\newcommand{\pa}{\partial}
\newcommand{\Nzero}{\mathbb{N}_{\geq 0}}

\begin{document}

\title{Casimir self-energy of a $\boldsymbol{\delta}$-$\boldsymbol{\delta'}$ sphere}
\author{C. Romaniega}\email{cesar.romaniega@uva.es}\affiliation{Departamento de F\'isica Te\'orica, At\'omica y \'Optica,Universidad de Valladolid,Valladolid, 47011, Spain.}
\author{J. M. Munoz-Castaneda}\email{jose.munoz.castaneda@uva.es}\affiliation{Departamento de F\'isica Te\'orica, At\'omica y \'Optica,Universidad de Valladolid,
  Valladolid, 47011, Spain.} 
\author{I. Cavero-Pel\'aez}\email{cavero@unizar.es.com}\affiliation{Departamento de F\'isica Te\'orica, Facultad de Ciencias, Universidad de Zaragoza, Zaragoza, 50009, Spain}

\begin{abstract}
We extend previous work on the vacuum energy of a massless scalar field in the presence of singular potentials. We consider a single sphere defined by the so-called $\delta$-$\delta'$ interaction. Contrary to the Dirac $\delta$-potential, we find a nontrivial one-parameter family of potentials such that the regularization procedure gives an unambiguous result for the Casimir self-energy. The procedure employed is based on the zeta function regularization and the cancellation of the heat kernel coefficient $a_2$. The results obtained are in agreement with particular cases, such as the Dirac $\delta$ or Robin and Dirichlet boundary conditions.

\end{abstract}
\maketitle
\section{Introduction}
Quantum vacuum fluctuations are known to give rise to forces between two distinct bodies as well as pressure on the surface of a single object. This macroscopic manifestation of the vacuum state associated to quantum fields has been investigated and measured in some special cases, achieving a level of concordance between theory and experiments that has astonished the community
(see Refs. \cite{bordag2009advances,milton2009zeropoint} for general reviews). From a quantum field theoretical point of view, the zero point energy due to the quantum vacuum fluctuations carries divergences. The appearance of the Casimir energy has stressed the importance of dealing with divergences and acquiring a deep understanding of their nature to the point of extracting the finite part of the zero point fluctuations, isolating the different divergent contributions and obtaining a physically meaningful result. After regularization and renormalization, the part of the quantum vacuum energy that encloses the quantum vacuum interaction between two objects is in general unambiguous and leads to a finite force between the bodies \cite{bordag2009advances,kenneth2006opposites,kenneth2008casimir,rahi2009scattering}. However, in general, the self-energy of a single object is only unambiguously defined for the case of massive quantum vacuum fluctuations. In the case of massless quantum fields the self-energy is only defined in a unique way {for a few} cases involving special geometries and boundary conditions. 
For example, it is well known that in the dilute case the Casimir energy, that can also be calculated as the sum of the van der Waals interactions, is unambiguously identified once the surface and volume divergences are removed \cite{PhysRevLett.82.3948, Barton_1999,ROMEO2005309,marachevsky2001casimir}. Perfectly conducting {bodies}, as well as dielectric geometries such as spheres or cylinders have been computed resulting on finite answers for the Casimir stress on the surface \cite{boyer1968quantum, deraad1981casimir, PhysRevE.55.4207, cavero2005casimir}

In all the cases mentioned above, different techniques for regularizing the vacuum self-stress and extracting the divergences have been used. From the $\zeta$-function regularization, to point splitting, analytic continuation or the calculation of heat kernel coefficients, several methods are used to understand the meaning and nature of the infinities arising from summing the frequencies of vacuum fluctuations of the zero-point. The difficulties that the study of the self-energies carries have been discussed broadly, in particular in spheres with a singular potential. 
Bordag et al. were the first ones discussing these divergences by computing the heat kernel coefficients \cite{bordag1999ground}. Furthermore, Bordag, Kirsten, Vassilevich, and others, have given analytic formulas that enable the characterization of the infinities and the ambiguities appearing in the calculation of quantum vacuum self-energies in terms of the heat kernel coefficients \cite{kirsten2001spectral,vassilevich2003heat}.

In a recent paper \cite{cavero2021casimir}, we calculated the interaction energy between two concentric spherical shells mimicked by singular potentials of the type $\ddp$ on the surfaces. In this case the total vacuum energy can be written as
\begin{equation}
	E_0=E_1+E_2+E_\text{int},
\end{equation}
being $E_\text{int}$ the interaction energy, that we can calculate unambiguously, and $E_1$ and $E_2$ the self-energies of the first and second body respectively, that we need to study in order to check whether they are finite or they present irremovable ambiguities. Since the divergent contributions depend on the characteristics of the body, like the radius of the sphere, a renormalization procedure is needed in order to give a meaningful result.  Therefore, if the self-energy of each sphere separately is not uniquely defined and we consider concentric spheres, the total quantum vacuum energy will be, as well, not well-defined.

In this paper we focus on the pressure on a single sphere, thus extending the work of our previous paper \cite{cavero2021casimir}. In the latter, we studied the sign of the interaction energy for a massless scalar field in the presence of two concentric $\delta$-$\delta'$ spheres employing the TGTG representation of the energy.
The same representation was used in \cite{romaniega2021repulsive}, where the pressure acting on a dielectric sphere enclosed within a magnetodielectric cavity was studied. Although the sign of the interaction pressure was determined for quite general inhomogeneous permittivities and permeabilities, the self-pressure of the sphere was only well-defined in the known dilute limit \cite{bordag1999ground}. Indeed, there are a few cases in which the self-energy has an unambiguous meaning \cite{milton2008local}. One is the aforementioned dilute limit \cite{bordag1999ground,cavero2005casimir},  a magnetodielectric object when the speed of light is the same inside and outside {\cite{klich1999casimir,milton1999mode,brevik1982electrostriction,brevik1982casimir,brevik1985attractive,brevik1988casimir}} or a  perfectly conducting spherical or cylindrical shell \cite{boyer1968quantum,deraad1981casimir}. For a massless scalar field, an unambiguously finite result is found up to second order for the $\delta$-potential  weak limit \cite{milton2004casimirdelta}, as well as for Dirichlet and Neumann boundary conditions. For these boundary conditions a cancellation of the divergences occurs when the whole space is considered \cite{bordag2009advances}. However, for Robin boundary conditions this is no longer the case, and the cancellation only occurs for certain values of the parameter \cite{kirsten2001spectral}.

{In this paper we} employ the zeta function regularization for {in order to analyze} the divergences. Within this approach, the energy is expressed in terms of the  zeta function associated with a Schr\"odinger-type operator $P$
\begin{equation}\label{eq:Zeta1}
	E_{0}(s)=\frac{{\mu}^{2s}}{2} ~ \sum_{n}\omega_n^{1-2s}=\frac{{\mu}^{2s}}{2} ~ \zeta_{P}(s-\frac 1 2 ),
\end{equation}
where $\mu$ is a parameter with dimensions of mass introduced to keep the right dimensions and $\hbar=c=1$. The zeta function associated with the operator determining the modes of the system is 
\begin{equation}\label{eq:Zeta2}
	\zeta_{P}(s)=\sum_n \lambda_{n}^{-s},\quad
	P \varphi_{n}(\boldsymbol{x})=\lambda_{n} \varphi_{n}(\boldsymbol{x}).
\end{equation}
In our case, we have $P=-\Delta+	V_{\delta\text{-}\delta'}(r)$, where the potential represents a spherical singular interaction. From the asymptotic behaviour of the eigenvalues of this operator, indeed, for any  second order elliptic differential operator \cite{weyl1912asymptotische},
the sum \eqref{eq:Zeta1} is divergent for $s=0$ and {it needs} to be regularized to find a meaningful result. Once the divergences are identified, we need to renormalize the resulting expression.  Bordag, Kirsten, Vassilevich and their collaborators demonstrated that the self-energy for massless scalar fields is defined in a unique way only if the heat kernel coefficient $a_2$ of the operator $P$ given above is identically zero.

The aim of this paper is to study  a complicated enough interaction to obtain nontrivial systems for which $a_2=0$, unlike what happens for the $\delta$-potential, and simple enough to proceed in an analytic way. The $\delta$-$\delta'$ potential is chosen since it has two couplings that will enable, for certain particular cases, the cancellation of the $a_2$ heat kernel coefficient. This point interaction was introduced in \cite{kurasov1996distribution} and studied in different contexts over the years \cite{gadella2009bound,munoz2015delta,albeverio2000singular,martin2022solvable}, where many analytical results have been obtained.

The paper is organized as follows. In section \ref{sec-ddp} we show basic previous results concerning the $\ddp$ potential concentrated in a spherical shell obtained in \cite{munoz2019hyperspherical}. In Section \ref{sec:Zeta} we compute the quantum vacuum energy for a three-dimensional spherical shell mimicked by a radial $\ddp$ potential using the zeta function regularization, and obtain those particular values for the couplings that give rise to a heat kernel coefficient $a_2$ identically zero. Section \ref{sec:ecas} shows the numerical results for the finite quantum vacuum self-energy and pressure {when it is unambiguously} defined ($a_2=0$).  Finally, in Section \ref{sec:conclu} we present our conclusions and further comments. At the end, we include Appendix A and B where we present the derivation of the matching conditions and Jost function of the $\ddp$ interaction as well as some particular cases that enable us to check our calculations by obtaining results already published by other authors.

\section{$\boldsymbol{\delta}$-$\boldsymbol{\delta'}$ potential on a spherical shell}\label{sec-ddp}
Let us consider a single spherical shell defined by the singular potential
\begin{equation}\label{eq:potU}
	V_{\delta\text{-}\delta'}(r)=\lambda_{0} \hspace{0mm}\delta  (r-r_0)+2 \lambda_{1}\hspace{0mm}\delta' (r-r_0),\quad r_0\in\mathbb{R}^+.
\end{equation}
The system of units chosen implies that
$[\lambda_{0}]=L^{-1},$ and  $[\lambda_{1}]= 1.$
The scalar field satisfies the Klein-Gordon equation which, after taking its time Fourier transform, is
 \begin{equation}\label{eq:schrH}
 	\left[-\Delta+V_{\delta\text{-}\delta'}(r)\right]\varphi(\mathbf{x}) = \omega^2 \varphi(\mathbf{x}).
 \end{equation}
Due to the spherical symmetry of the system, the solutions can be written as
\begin{equation}\label{solution_decomposition}
	\varphi(\mathbf{x}) =\sum_{\ell=0}^\infty \sum_{m=-\ell}^\ell \rho_\ell(r)Y_{\ell m}(\theta,\phi),
\end{equation}
where $Y_{\ell m}(\theta,\phi)$ are the spherical harmonics.
The non-relativistic Schr\"odinger Hamiltonian in Eq.~\eqref{eq:schrH} has been studied in detail in \cite{munoz2019hyperspherical}, where the potential $V_{\delta\text{-}\delta'}(r)$ is defined by matching conditions on the surface of the sphere with radius $r_0$ over the space of field modes as 
\begin{equation}\label{eq:matchG}
	\left(
	\begin{array}{c}
		\rho_\ell(r_0^+) \\
		\rho'_\ell(r_0^+) \\
	\end{array}
	\right)=\left(
	\begin{array}{cc}
		\alpha  & 0 \\
		\widetilde{\beta} & {\alpha^{-1} } \\
	\end{array}
	\right)\left(
	\begin{array}{c}
		\rho_\ell(r_0^-) \\
		\rho'_\ell(r_0^-) \\
	\end{array}
	\right).
\end{equation}
The prime here, and  throughout the text, indicates derivative with respect to the argument and
$r_0^{\pm}$ denotes the limit to $r_0$ taken from the right or from the left, respectively.
 We have also defined 
\begin{equation}\label{eq:defs}
	\alpha= \frac{1+\lambda_1}{1-\lambda_1},\quad\widetilde\beta= \frac{\widetilde \lambda_{0}}{1- \lambda_{1}^2},\quad \widetilde \lambda_{0}=-\frac{4\hspace{0mm} \lambda_{1}}{r_0}+{\lambda_{0}}.
\end{equation}
In Appendix~\ref{appendix:A} we show a derivation of the matching conditions \eqref{eq:matchG}.
These  conditions are ill-defined if $\lambda_1=\pm1$. In these cases we can write \cite{albeverio2000singular}
\begin{equation}\label{eq:RobinDirichlet}
	\begin{aligned}
		&\rho_{\ell}(r_0^-)=0, \quad \rho'_{\ell}(r_0^+)-D^+\rho_{\ell}(r_0^+)=0 \quad\text{if}\quad \lambda_1=+1,\\
		& \rho_{\ell}(r_0^+)=0,\quad\rho'_{\ell}(r_0^-)+D^-\rho_{\ell}(r_0^-)=0\quad\text{if}\quad \lambda_1=-1,
	\end{aligned}
\end{equation}
where $D^\pm=4/(\lambda_0\mp 4r_0^{-1})$. (Notice that there is a typo in Eq.(30) in \cite{cavero2021casimir}).  

The eigenvalues are not known for this problem, so the explicit summation shown in \eqref{eq:Zeta2} can not be performed. However, in \cite{munoz2019hyperspherical} the scattering problem for the radial $\ddp$ potential was solved. Therefore using Cauchy's formula, we can employ the following expression to study the Casimir self-energy, 
\begin{equation}\label{eq:E0}
 {E_{0}(s)}=-\mu^{2s}{\cos \pi s\over \pi}\sum\limits_{\ell=0}^{\infty}\nu
\int\limits_{0}^{\infty}\mathrm{d} \kappa \ \kappa^{1-2s}{\pa\over\pa \kappa}\log
f_{\ell}(\kappa),
\end{equation}
in terms of the Jost function $f_{\ell}(\kappa)$, where the volume energy has already been subtracted \cite{bordag1996vacuum}.
For each value of the angular momentum this function satisfies \cite{taylor2006scattering}
\begin{equation}\label{eq:fJostPhase}
\frac{f_\ell(\omega)}{f_\ell^*(\omega)}=e^{-2 i \delta_\ell(\omega)},
\end{equation}
being $\delta_\ell(\omega)$ the scattering phase shift. The latter has been computed  in \cite{munoz2019hyperspherical} for the potential \eqref{eq:potU}. In Appendix~\ref{appendix:A} we prove that the Jost function can be written as
\begin{equation}\label{jost}
	\begin{aligned}
f_\ell(\kappa)\!=\!1+\!\dfrac{\lambda_0 r_0\!-\!2 \lambda_1 }{\lambda_1^2+1}I_{\nu}(y ) K_{\nu}(y )
\!-\!\dfrac{2 \lambda_1 y }{\lambda_1^2+1}(I_{\nu}(y ) K_{\nu}(y ))'\!,
	\end{aligned}
\end{equation}
where it is assumed that
\begin{equation}
	\omega= i \kappa, \quad \nu=\ell+\frac{1}{2},  \quad {y}=\kappa {\it r}_{{0}}.
\end{equation}
 By turning off the coefficient of the $\delta'$ term in the potential, this expression reduces to the known result of the Jost function corresponding to a spherical shell with a $\delta$-potential on its surface \cite{bordag1999ground},
 \begin{eqnarray}\label{eq:Jostdelta}
 	f_\ell(\kappa)&=&1+r_0 \lambda_0 I_{\nu}(y) K_{\nu}(y).
 \end{eqnarray}

\section{Zeta function regularization}\label{sec:Zeta}

The zeta function is connected with the heat kernel $K(t)$ through the Mellin transform
\begin{equation}\label{eq:Zeta3}
\zeta (s)={1\over \Gamma (s)}\int_{0}^{\infty}t^{s-1} K(t)\,\mathrm{d} t ,
\end{equation}
where
\begin{equation}
K(t) = \sum_n e^{-\lambda_{n} t}.
\end{equation}
At $s=0$ this expression is exponentially decreasing for large $t$. The trouble comes when $t$ is small. For that we use the asymptotic expansion  \cite{vassilevich2003heat}
\begin{equation}\label{eq:Kt}
K(t)\sim {1\over (4\pi t)^{3/2}}\sum\limits_{n}a_{n/2}t^{n/2}.
\end{equation}
Taking these expressions into account and computing the Casimir energy for massless scalar field as in Eq.~\eqref{eq:Zeta1}, we find the result
\begin{equation}\label{eq:E0as2}
E_0^\text{as}(s) = - \dfrac{a_2}{32 \pi ^2}\left(\dfrac{1}{s}+2\log (\mu r_0)\right) + E_0^\text{an} + O(s),
\end{equation}
where the superindex -as stands for asymptotic and -an for the analytic part. The presence of the log term does not allow us to prescribe a proper renormalization procedure without ambiguities for massless quantum fluctuations\footnote{For the case of massive quantum fluctuations with mass $m$ the condition $E_0^\text{ren}(m\to\infty)=0$, ensures that there are no ambiguities.}. The only way to get a universal answer for the Casimir self-stress is to make sure this term is not present in the computation of the energy. Therefore, we look for those cases where $a_2=0$ which can be identified as the coefficient of the divergent term in the vacuum energy since $s\rightarrow 0$.

The coefficient $a_2$ can be computed analyzing the behavior of the zeta function given in Eq.~\eqref{eq:Zeta1} if we take into account Eqs.~\eqref{eq:Zeta3} and \eqref{eq:Kt}. In particular,
\begin{equation}\label{eq:HeatKernelRes}
	a_{n}=\underset{s=\frac{3}{2}-n}{\text{Res}}((4\pi)^{\frac{3}{2}}\Gamma \left(s\right)\zeta(s)),\quad 2n\in \Nzero. 
\end{equation}
In our particular case, the self-energy of our configuration is given by the expression in Eq.~\eqref{eq:E0} which is not well-defined for $s=0$.
Performing an analytic continuation of the function $E_0(s)$ in $s$, enables the identification of the divergence of the vacuum energy as a simple pole at $s=0$.
To do so, we subtract and add the asymptotic behavior of the Jost function and define
\begin{equation}\label{eq:E0fin}
E_0^\text{fin}=-{1\over \pi}\sum\limits_{\ell=0}^{\infty}\nu\int\limits_{0}^{\infty}\mathrm{d} \kappa \ \kappa{\pa\over\pa \kappa}\left(\log
f_{\ell}(\kappa)-\log f^\text{as}_\ell(\kappa)\right),
\end{equation}
as the finite part of the energy at $s=0$ and
\begin{equation}\label{eq:E0as}
	E_0^\text{as}(s)=-\mu^{2s}{\cos \pi s\over \pi}\sum\limits_{\ell=0}^{\infty}\nu\int\limits_{0}^{\infty}\mathrm{d} \kappa \ \kappa^{1-2s}{\pa\over\pa \kappa}\log f^\text{as}_\ell(\kappa),
\end{equation}
the asymptotic one. The study of the latter at $s=0$ gives the pole that corresponds to the $a_2$ coefficient. Since the main contribution comes from large $\ell$, we use the  uniform asymptotic expansion of the modified Bessel functions (see for example \cite{olver2010nist}) in Eq.~\eqref{jost} where $y\equiv z\nu$.

The number of terms we subtract in Eq.~\eqref{eq:E0fin} is determined by requiring this quantity to become finite. We achieve that by expanding the Jost function up to third order in $1/\nu$ making use of the mentioned uniform asymptotic expansion.  This allows us to write the argument of the logarithm as
\begin{equation*}
    f^\text{as}_\ell(\kappa)\approx 1+x(\nu),
\end{equation*} 
where $x(\nu)$ is a function of $\nu$ such that $x(\nu)\rightarrow0$ when $\nu\rightarrow\infty$. Then, we expand the logarithm in Eq.~\eqref{eq:E0as} as a power series,
\begin{equation}\label{eq:logfas}
\log f^\text{as}_\ell(\kappa)\equiv\log f^\text{as}_\ell(z)=\sum_{n=1}^{N=3} \sum_{i=n}^{3n} C_{n,i}\frac{t^i(z)}{\nu^n},
\end{equation}
where $t=1/\sqrt{1+z^2}$ and $z = \kappa r_0/\nu$. Subtracting the first three terms ($N=3$) of the asymptotic expansion is enough to isolate the analytic part of the zeta function. The first nonzero coefficients $C_{n,i}$ are

\begin{widetext}
\begin{eqnarray*}
	&C_{1,1}&=\dfrac{\lambda_0 r_0}{2 (\lambda_1^2+1)}, \  C_{1,3}=-\dfrac{\lambda_1}{\lambda_1^2+1},\\[0.5ex]
&	C_{2,2}&=-\dfrac{\lambda_0^2 r_0^2}{8 \left(\lambda_1^2+1\right)^2}, \  C_{2,4}=\dfrac{\lambda_0 \lambda_1 r_0}{2 \left(\lambda_1^2+1\right)^2}, \  C_{2,6}=-\dfrac{\lambda_1^2}{2 \left(\lambda_1^2+1\right)^2},\\[0.5ex]
&	C_{3,3}&=\dfrac{2 \lambda_0^3 r_0^3+3 \left(\lambda_1^2+1\right)^2 (4 \lambda_1+\lambda_0 r_0)}{48 \left(\lambda_1^2+1\right)^3}, \  C_{3,5}=-\dfrac{2 \lambda_0^2 \lambda_1 r_0^2+3 \left(\lambda_1^2+1\right)^2 (9 \lambda_1+\lambda_0 r_0)}{8 \left(\lambda_1^2+1\right)^3}, \\[0.5ex]
& C_{3,7}&=\dfrac{120 \lambda_1 \left(\lambda_1^2+1\right)^2+\lambda_0 \left(5 \lambda_1^4+18 \lambda_1^2+5\right) r_0}{16 \left(\lambda_1^2+1\right)^3}, \  C_{3,9}=-\dfrac{\lambda_1 \left(105 \lambda_1^4+218 \lambda_1^2+105\right)}{24 \left(\lambda_1^2+1\right)^3}.
\end{eqnarray*}
\end{widetext}
 For $\lambda_1=0$, i.e., $\delta$-potential, we obtain the coefficients $X_{n,i}$ found in  \cite{bordag1999heat} and \cite{bordag1999ground}, although  there is a minus sign missing in the coefficient $X_{2,2}$ in \cite{bordag1999ground}. 
\subsection{Heat kernel coefficient $a_2$}
The purpose of this section is to analyze  $E_0(s)$ as $s\to 0$ in order to discuss the divergences. The relevant term is $E_0^\text{as}(s)$ since $E_0^\text{fin}$ is analytic at $s=0$ and gives no contribution to the residues of $E_0(s)$.

The integral to be computed in  \eqref{eq:E0as} after performing  the change of variables $z=\kappa r_0/\nu$ is
\begin{equation*}
I=\frac{(r_0\mu)^{2s} }{r_0}{\nu }^{1-2 s}\int\limits_{0}^{\infty}\mathrm{d} z \ z^{1-2s}{\pa\over\pa z}\log f^\text{as}_\ell(z),
\end{equation*}
which can be easily solved using
\begin{equation*}
\int_0^{\infty } \frac{z^n}{\left(z^2+1\right)^b} \, \mathrm{d}z=\frac{\Gamma \left(\frac{n+1}{2}\right) \Gamma \left(b-\frac{n}{2}-\frac{1}{2}\right)}{2 \Gamma (b)}.
\end{equation*}
Now we perform the sum over $\nu$ in $E_0^\text{as}(s)$. This can be written in terms of the Hurwitz zeta function for the three values of $n$
\begin{eqnarray*}
	\sum\limits_{\ell=0}^{\infty}\nu^{1-n+1-2s}=\sum\limits_{\ell=0}^{\infty}\nu^{2-n-2s}=\zeta \left(n+2s-2,\frac{1}{2}\right),
\end{eqnarray*}
that satisfies the following identity involving the Riemann zeta function
\begin{equation*}
\zeta \left(n+2s-2,\frac{1}{2}\right)=(2^{2s+n-2}-1)\zeta(n+2s-2).
\end{equation*}
From this zeta function, $n = 1$ gives a finite contribution and it vanishes for $n = 2$. It is the term $n=3$ of the zeta function the one that brings out a divergence. For $s\to 0$,
\begin{equation*}
    \zeta \left(1+2s,\frac{1}{2}\right)=\frac{1}{2 s}+(\gamma +2 \log 2)+O\left(s^1\right).
\end{equation*}where  $\gamma$ is Euler's constant \cite{olver2010nist}.
We substitute the above back in the integral shown in Eq.~\eqref{eq:E0as} and identify the $a_2$ coefficient of the singularity. We find
\begin{equation}\label{eq:a2}
a_2 = \dfrac{2 \pi  \left(128 \lambda_1^3+140 \lambda_0^2 \lambda_1 r_0^2-35 \lambda_0^3 r_0^3-224 \lambda_0 \lambda_1^2 r_0\right)}{105 \left(\lambda_1^2+1\right)^3 r_0},
\end{equation}
and the finite term in Eq.~\eqref{eq:E0as2} is given by
\begin{widetext}
\begin{eqnarray}\label{eq:E0an}
E_0^\text{an}&=&	\dfrac{1}{5040 \pi   r_0\left(\lambda_1^2+1\right)^{3}} \left(-64 \lambda_1^3 (12 \gamma -1+36 \log\! 2)-420 \lambda_0^2 \lambda_1 r_0^2 (2 \gamma -1+\log\! 64)+210 \lambda_0^3 r_0^3 (\gamma -1+\log\! 8) \right. \\ \nonumber
  &-&21 \left. \lambda_0 r_0 \left(5 \lambda_1^4 (-12 \log\! A+1+\log\! 8)-2 \lambda_1^2 (60 \log\! A-13+81 \log\! 2)-60 \log\! A+\gamma  \left(5 \lambda_1^4-54 \lambda_1^2+5\right)+5+15 \log\! 2\right)\right).
\end{eqnarray}
\end{widetext}
$A$ is the Glaisher's constant \cite{olver2010nist}. 
For $\lambda_1=0$, $\delta$-potential, we recover the results found in \cite{bordag1999heat,bordag1999ground}.
From \eqref{eq:a2}, we find that there is a family of parameters that make $a_2$ vanish and therefore defines the self-stress over the sphere without ambiguities. By making $a_2=0$ we find that
\begin{equation*}
	\begin{aligned}
		\lambda_1=\frac{1}{24} \left(\sqrt[3]{42 \sqrt{30}+224}-\frac{14^{2/3}}{\sqrt[3]{3 \sqrt{30}+16}}+14\right)\lambda_0 r_0.
	\end{aligned}
\end{equation*}
There are other two combinations of the couplings such that $a_2=0$, but they involve complex solutions.
Consequently, the self-energy is properly defined if
\begin{equation}\label{eq:c0}
	c_0\lambda_1= \lambda_0 r_0,\quad c_0\simeq 1.20818671192.
\end{equation} 

%
\section{Renormalized energy and pressure}\label{sec:ecas}
Once $a_2=0$, the renormalized energy is unambiguously defined
\begin{equation}\label{eq:E0ren}
E_0^\text{ren}=E_0^\text{fin}+E_0^\text{an}.
\end{equation}
In contrast to $E_0^\text{fin}$ and $E_0^\text{an}$, this quantity is uniquely defined since it does not depend on the number of terms subtracted.
As we have stated, the renormalization is completely determined if the heat kernel coefficient  $a_2=0$ \cite{bordag2009advances}.
For the $\delta$-potential this is only possible for the trivial case $\lambda_0=0$ \cite{bordag1999ground}, although the weak limit can be computed until second order \cite{milton2004casimirdelta}. Studying the divergences using Green's functions it is shown that they come from the surface term only \cite{cavero2006local}.

\subsection{Pressure on the sphere}

The pressure acting on the surface of the sphere can be obtained from $E_0^\text{ren}$.  To define this pressure we make use of the principle of virtual work. For our spherically symmetric system  \cite{barton2004casimir,li2019casimir}
\begin{equation}\label{eq:MeanPress}
	p^\text{ren}_0= -\dfrac{1}{4\pi r_0^{2}}\frac{\partial  E_0^\text{ren}}{\partial  r_0}.
\end{equation}
For the case in Eq.~\eqref{eq:c0}, where the self-energy is well-defined, the finite and analytic parts of the renormalized energy become
\begin{equation}\label{eq:E0fin2}
	E_0^\text{fin}={1\over \pi r_0}\sum\limits_{\ell=0}^{\infty}\nu^2\int\limits_{0}^{\infty}\mathrm{d} z\, (\log
	f_{\ell}(z)-\log f^\text{as}_\ell(z))
\end{equation}
and
\begin{equation}\label{eq:E0an2}
	E_0^\text{an}=\frac{F(\lambda_1,c_0)}{5040 \pi  \left(\lambda_1^2+1\right)^3 r_0},
\end{equation}
where the function $F(\lambda_1,c_0)$ is obtained from Eq.~\eqref{eq:E0an} when $\lambda_0r_0= \lambda_{1}c_0$. Both terms have the same dependence on $r_0$ and therefore,
\begin{equation}\label{eq:MeanPress}
	\text{sgn}\	p^\text{ren}_0= \text{sgn}\ E_0^\text{ren}.
\end{equation}
Indeed, $E_0^\text{ren}$ can be written as $E_0^\text{ren}={e_0^\text{ren}}/{r_0}$, being $e_0^\text{ren}$ a numerical constant independent of $r_0$. Then,
\begin{equation}
	E_0^\text{ren}=\frac{e_0^\text{ren}}{r_0},\quad p=\frac{e_0^\text{ren}}{4\pi r_0^4}.
\end{equation}
 
\subsection{Numerical evaluation}

Now we compute $E_0^\text{ren}$ when $a_2=0$. The finite part of the energy can be calculated only numerically and for that we use a code in the interpreted programming language Python.
We use the expression in Eq.~\eqref{eq:E0fin2}, noting that the boundary term vanishes for both zero and infinity. 

First, note that if \eqref{eq:c0} holds there are no bound states in the quantum mechanical sense. That is to say, with our potential in \eqref{eq:schrH} we can not have $\omega^2<0$. For the latter we should change the integration contour in order to avoid the poles along the imaginary axis \cite{taylor2006scattering}. The absence of bound states can be proved with Proposition 2 in \cite{munoz2019hyperspherical}. This result states that the quantum mechanical system admits bound states with angular momentum from 0 to $\ell_\text{max}$, being
\begin{equation}
\ell_\text{max} =\left\lfloor -\frac{1}{2}+\frac{\lambda_1-\frac{\lambda_0 r_0}{2}}{\lambda_1^2+1}\right\rfloor.
\end{equation}
For the $\delta$-potential case we deduce that there are no bound states unless the potential is deep enough, $\lambda_0 r_0<-1$, as expected. For the $\delta$-$\delta$' potential however, and under the condition in \eqref{eq:c0}, there exist no bound states regardless of the value of $\lambda_0$. This can be easily proved noting that in this case
\begin{equation}
\ell_\text{max} =-\frac{1}{2}\left\lfloor1+ (2-c_0)\frac{\lambda_1}{\lambda^2_1+1}\right\rfloor,\quad \lambda_1= \frac{\lambda_0 r_0}{c_0},
\end{equation}
which is always negative since $c_0\in(0,2)$. Indeed, this is clear if $\lambda_1\leq0$.  If $\lambda_1>0$ we have $\lambda_1/\left(\lambda_1^2+1\right)\leq1/2$.
Note that for $\lambda_1=1/2$, where $\ell_\text{max}=0$ there is no zero-mode either, see Sec. 4.2 of \cite{munoz2019hyperspherical}.
 In addition, from \cite{munoz2019hyperspherical} we know that bound states for positive values of $\lambda_0$ are only possible for two spatial dimensions.
 
\begin{figure}[h!]
	\centering
	\includegraphics[width=0.48\textwidth]{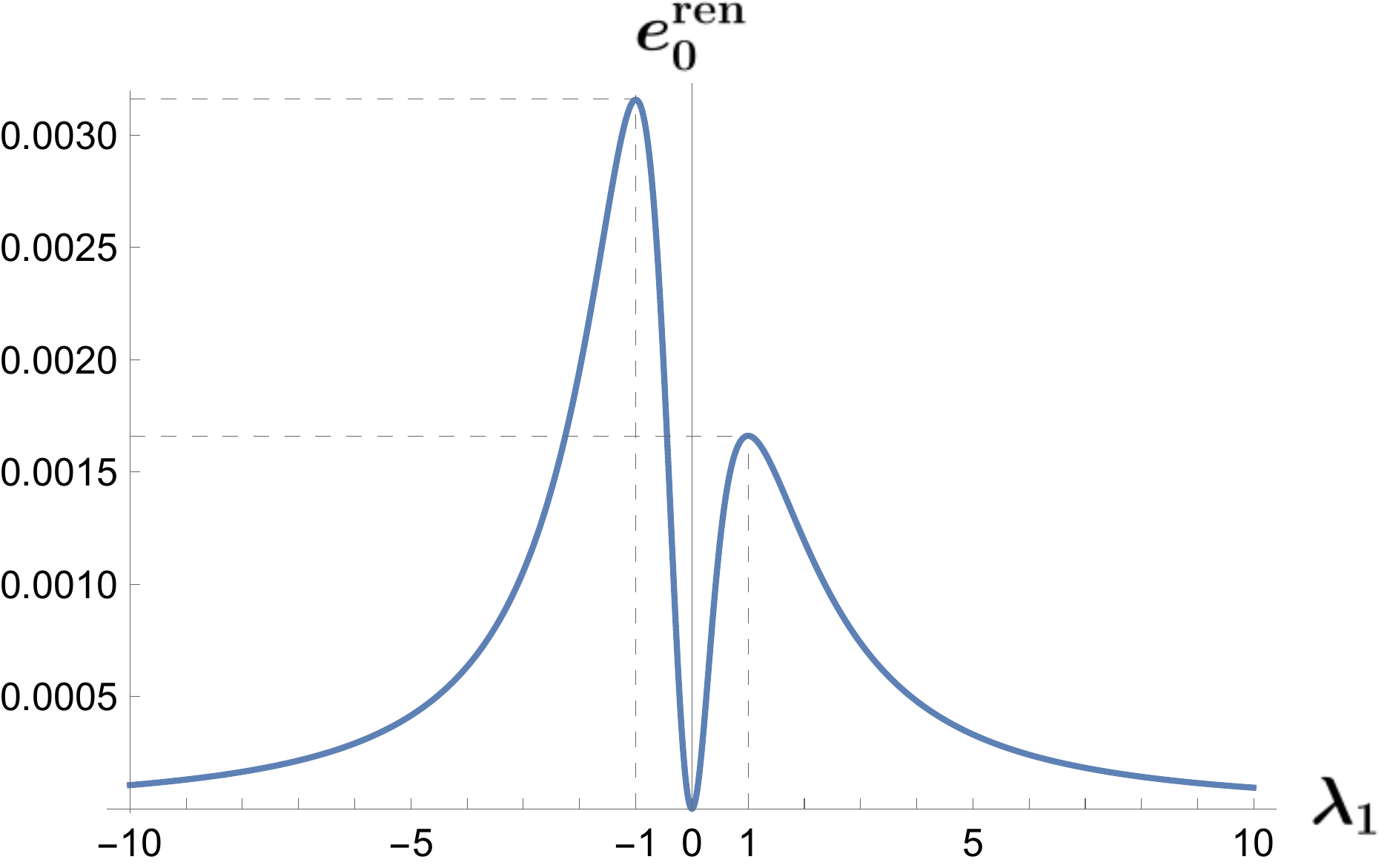}
	\caption{Renormalized energy \eqref{eq:E0ren} for $r_0=1$, $E_0^\text{ren}=e_0^\text{ren}/r_0$, and $a_2=0$, i.e., $c_0\lambda_1=\lambda_0$. } 
	\label{fig:E_Ren_a2_zero}
\end{figure}

We plot the renormalized vacuum energy \eqref{eq:E0ren} as a function of $\lambda_{1}$, Fig.~\ref{fig:E_Ren_a2_zero}, using Eqs.~\eqref{eq:E0fin2} and \eqref{eq:E0an2}. Note that only positive values are obtained, i.e., self-repulsion which tends to expand the sphere.
A similar behaviour for the energy was found for the interaction energy between two concentric $\delta$-$\delta'$ spheres \cite{cavero2021casimir}.
For instance, the result is not symmetric under the change $\lambda_{1}\to-\lambda_{1}$, contrary to what happens in the $\delta$-$\delta'$ plates \cite{munoz2015delta}. In addition, the maximum values of the energy are found when we approach $\lambda_1=\pm 1$ corresponding to Dirichlet and Robin boundary conditions, see Eq.~\eqref{eq:RobinDirichlet}. In particular, we have
\begin{equation*}
	\begin{aligned}
		&\rho_{\ell}(r_0^-)=0, \quad \rho'_{\ell}(r_0^+)+\frac{4 r_0}{4-c_0}\rho_{\ell}(r_0^+)=0 \quad\text{if}\quad \lambda_1=+1,\\
		& \rho_{\ell}(r_0^+)=0,\quad\rho'_{\ell}(r_0^-)+\frac{4 r_0}{4+c_0}\rho_{\ell}(r_0^-)=0\quad\text{if}\quad \lambda_1=-1.
	\end{aligned}
\end{equation*}
Consequently, we have found a system, combination of Dirichlet and Robin boundary conditions (inside and outside the sphere) whose renormalized self-energy is well-defined. Notice that this is a non-trivial result since we know that for Dirichlet or Neumann boundary conditions $a_2\neq 0$ when only the interior or exterior region is considered \cite{bordag2009advances}. Moreover, since they only depend on odd powers of the extrinsic curvature, when the interior and exterior of the sphere are considered, the divergences cancel each other. For Robin boundary conditions even powers are also present, and the cancellation only occurs for special values of the Robin parameter \cite{kirsten2001spectral}.  

Furthermore, the value of the self-energy and the self-pressure are of the same order of magnitude that the one found for Dirichlet boundary conditions, where $E_0^\text{ren}\simeq 0.0028168/r_0$. However, the result presented here is significantly lower than the one found for Neumann boundary conditions, where $E_0^\text{ren}\simeq -0.2238216/r_0$. 
It is worth noting that $\ddp$ interaction imposes Dirichlet boundary conditions in the so-called strong limit: $\lambda_{0}\to\infty$ and $\lambda_{1}=0$. However, although we have Neumann boundary conditions in the limits $\lambda_1\to\pm 1$ and $\lambda_0\to\infty$, we can not impose these boundary conditions at both sides of the spherical shell with the $\ddp$ interaction \eqref{eq:RobinDirichlet}.

In the limit $\lambda_1 \to \pm\infty$ the self-energy goes to zero. This is in agreement with previous results
\cite{romaniega2020approximation} since this case corresponds to the case with no potential (notice that the fields and their derivatives become continuous at the boundary, Eq.~\eqref{eq:matchG}).

In Appendix~\ref{appendix:B}, we include consistency checks of our calculations. To do so, we have obtained the weak limit formula first computed in \cite{milton2004casimirdelta} for a Dirac $\delta$ spherical shell, and reproduced the numerical results in \cite{leseduarte1996complete} for a spherical shell in the decoupled limit ($\lambda_1\to\pm 1$).

\section{Conclusions}\label{sec:conclu}

In this paper we have added another example to the short list of simple configurations in which the Casimir self-energy for a massless field is unambiguously defined. This occurs due to a particular cancellation between the Dirac $\delta$ and the $\delta'$ interaction, Eq.~\eqref{eq:c0}. A similar  cancellation arises when considering the Dirac $\delta$ and other type of singular interaction defined by imposing matching conditions such that the derivative is continuous and there is a finite discontinuity in the radial function \cite{milton2004casimir}. 

For the one-parameter family of values in which the energy and pressure are well-defined we only find positive values of both quantities. This leads to self-repulsion which tends to expand the sphere. The first example of self-repulsion was found by Boyer \cite{boyer1968quantum}, ruling out the idea that the Casimir energy could stabilize the electron.

In Appendix~\ref{appendix:B} we have tested our results with the Dirac $\delta$ weak limit and Robin-Dirchlet boundary conditions. It is worth mentioning that our values of energy and pressure are similar to the ones obtained for a Dirichlet sphere. In addition, from this approach we see that we do not need to consider the interior and the exterior in an independent way. This is clear for matching conditions, but we also obtain this in the limit when we approach boundary conditions, even though a Jost function only sees the exterior region for any opaque potential.


\section*{Acknowledgments}

C.R. is grateful to G. Fucci for fruitful discussions  at East Carolina University. C.R. thanks the Spanish Government for funding under the FPU-fellowships program FPU17/01475 and the FPU mobility program  EST21/00286.  C.R. acknowledges the support of the grant PID2020-113406GB-I0 funded by the MCIN (Spanish Government). I.C.P. would like to thank the support received from the grants PGC2018-095328-B-I00 funded by the {\it Agencia Estatal de Investigación} (Spanish Government) and the FEDER fund (EU), and 225351 funded by the DGA. The authors would like to thank the valuable comments received from the referee.

\appendix
\section[{Singular $\ddp$ interaction}]{Singular $\boldmath\ddp$ interaction in 3D}\label{appendix:A}
\subsection{Calculation of the matching conditions for the 3D spherical shell}

In this appendix we will obtain the matching conditions that define the $\ddp$ spherical shell. From \eqref{eq:schrH} the Schrödinger Hamiltonian that characterizes the one-particle of the quantum vacuum fluctuations is given by
\begin{equation}
	H_{\ddp}=-\Delta+V_{\ddp}(r).
\end{equation}
Since the potential depends only on the radial coordinate, the previous Hamiltonian is spherically symmetric. Hence the eigenvalue problem
\begin{equation}\label{eq-a2}
	\left[-\Delta+V_{\ddp}(r)\right]\psi(\mathbf{x}) = E \psi(\mathbf{x}),
\end{equation}
is separable in spherical coordinates. Writing\footnote{As usual, $Y_{\ell,m}(\Omega)$ are the spherical harmonics, and $\Omega=(\theta,\phi)$ are the usual spherical angular coordinates.} $\psi(\mathbf{x})=\rho_\ell(r)Y_{\ell,m}(\Omega)$ Eq.~\eqref{eq-a2} becomes
\begin{equation}\label{eqI:difR}   
	\left[- \partial_r^2-\!\dfrac{2}{r}\partial_r+ \dfrac{\ell(\ell+1)}{r^2}+V_{\ddp} \right]{\rho}_{\ell}(r)=E \rho_{\ell}(r).
\end{equation} 
Now, introducing the reduced radial function 
$$u_{\ell}(r)=r {\rho}_{\ell}(r)$$
we end up obtaining a one-dimensional Schrödinger Hamiltoninan over the semi-axis $r\in (0,\infty)$ for each angular momentum
\begin{equation}
	H_\ell=-\partial_r^2+ \dfrac{\ell(\ell+1)}{r^2}+V_{\ddp}(r).
\end{equation} 
Finally, equation \eqref{eq-a2} can be written for each angular momentum $\ell$ in terms of the reduced radial function as
\begin{equation}
	H_\ell u_\ell(r)=E u_\ell(r),
\end{equation}
which is nothing but a collection of non-relativistic one-dimensional Hamiltonians over the semi-axis\footnote{The one-dimensional Hamiltonian $H_\ell^{(0)}=-\partial_r^2+\ell(\ell+1)r^{-2}$ over the semi-axis $(0,\infty)$ is essentially self-adjoint for any $\ell>0$. For the case  $\ell=0$ we choose the self-adjoint extension that maintains the scale invariance (see Appendix A in Ref. \cite{santagata}).} $(0,\infty)$. Since the potential in \eqref{eq-a2} depends on the radial coordinate $r$ our problem has spherical symmetry. As a consequence of the spherical symmetry, the matching conditions that define $V_{\ddp}$ in each $H_\ell$ must be the same for any value of $\ell$, otherwise the spherical symmetry would not hold (see Ref. \cite{romaniega2020approximation}). In particular, when $\ell=0$ our reduced Hamiltoninan becomes 
\begin{equation}\label{eqI:Pw}
	H_{\ell=0}=-\partial_r^2+ \lambda_0\delta(r-r_0)+2\lambda_1\delta'(r-r_0).
\end{equation}
This Hamiltonian has been previously studied in literature. The most rigorous study was done by Kurasov in Ref.~\cite{kurasov1996distribution}, where the previous one-dimensional Hamiltoninan is considered over the whole real line. For the sake of simplicity, and without loss of generality we can consider the change of variable $x=r-r_0$, and write
\begin{equation}\label{eqI:P}
	H_{\ell=0}=-\partial_x^2+ \lambda_0\delta(x)+2\lambda_1\delta'(x),
\end{equation}
where now $x\in(-r_0,0)\cup(0,\infty)$. To follow Kurasov's work we must assume that the $\delta'$-term in Eq.~\eqref{eqI:P} is the \textit{generalized derivative} of the Dirac-$\delta$ in the distributional sense. The term generalized derivative is used since we necessarily have discontinuous wave functions and the standard theory of distributions can not be applied. In this sense, it is convenient to define
 \begin{equation*}
		\bar{u}(0)=\dfrac{u(0^+)+u(0^-)}{2}. 
	\end{equation*}
where $u(0^+)$ and $u(0^-)$ are the values of the reduced radial function evaluated at zero when this point is approached from the positive and negative axes respectively. 
 In order to find the matching conditions at $r=r_0$ ($x=0$), we integrate, in the neighborhood of $r_0$, the differential equation $H_{\ell=0} u(x)=E u(x)$ between $-\varepsilon$ and $\varepsilon$ and then we make the limit $\varepsilon\rightarrow 0$.  With $H_{\ell=0}$ given in  \eqref{eqI:P} we get,
 \begin{eqnarray}\label{eq:Int1}
		 &-&\int_{-\varepsilon}^{\varepsilon}\frac{\rm d^2}{{\rm d} x^2}u(x){\rm d}x +\lambda_0\int_{-\varepsilon}^{\varepsilon}\delta(x)u(x){\rm d}x+\nonumber\\
   &&2\lambda_1\int_{-\varepsilon}^{\varepsilon}\Big(\frac{\rm d}{{\rm d} x}\delta(x)\Big)u(x){\rm d}x=\int_{-\varepsilon}^{\varepsilon} E u(x){\rm d}x,
	\end{eqnarray}
 The function $u(x)$ has a discontinuity at $x=0$ ($r=r_0$) that corresponds to the radius of the sphere, and therefore the integral in the second term of the expression above may not be well defined. However, since the values $u(0^+)$ and $u(0^-)$ are well defined and finite, we adopt an average prescription (previously used for example in Refs.~\cite{griffiths1993boundary,cavero2008NonConntG}) and take 
 \begin{equation*}
     \int_{-\varepsilon}^{\varepsilon}\delta(x)u(x){\rm d}x=\bar{u}(0).
 \end{equation*}
 In the third term we use integration by parts, the boundary term vanishes and we are left with 
 \begin{equation*}
     -\int_{-\varepsilon}^{\varepsilon}\delta(x)\frac{{\rm d}u(x)}{{\rm d} x}{\rm d}x=-\bar{u}'(0).
 \end{equation*}
 Grouping all terms together in \eqref{eq:Int1} we find a discontinuity in the first derivative,
 \begin{equation}\label{eq:Matching1}
     u'(0^+)-u'(0^-)=\frac{\lambda_0}{2}\big(u(0^+)+u(0^-)\big)-\lambda_1\big(u'(0^+)+u'(0^-)\big)
 \end{equation}
 The condition for the finite jump discontinuity of the function at zero is obtained by integrating twice the same differential equation. Considering $x'$ a variable in the same domain as $x$  we write
 \begin{equation*}
     \int_{-\varepsilon}^{\varepsilon}\int_{-\varepsilon}^{x}H_{\ell=0}\, u(x')\,{{\rm d} x'}{{\rm d} x}=-\int_{-\varepsilon}^{\varepsilon}\int_{-\varepsilon}^{x}E \,u(x')\,{{\rm d} x'}{\rm d}x
 \end{equation*}
 We follow the same procedure taking into account that now we can not throw the boundary term. Instead, it gives a contribution to the discontinuity in the radial function,
 \begin{eqnarray}
     &2&\lambda_1\int_{-\varepsilon}^{\varepsilon}\int_{-\varepsilon}^{x}\frac{\rm d}{{\rm d} x'}\big(\delta(x')u(x')\big){{\rm d} x'}{\rm d}x=\nonumber\\
     &=&2\lambda_1\int_{-\varepsilon}^{\varepsilon}\delta(x)u(x){\rm d}x= 2\lambda_1\bar{u}(0).
 \end{eqnarray}
 The other term different from zero comes from the second derivative in the Hamiltonian. As a consequence we get,
 \begin{equation*}
     -\Big(u(0^+)-u(0^-)\Big)+\lambda_1\Big(u(0^+)+u(0^-)\Big)=0,
 \end{equation*}
 giving rise to one of the equations we were looking for,
 \begin{equation}\label{eq:Matching2}
     u(0^+)= \frac{1+\lambda_1}{1-\lambda_1}\,u(0^-).
 \end{equation}
 Inserting this result in \eqref{eq:Matching1} and grouping terms we find,
 \begin{equation}\label{eq:Matching11}
     u'(0^+)=\frac{\lambda_0}{1-\lambda_1^2}\,u(0^-)+ \frac{1-\lambda_1}{1+\lambda_1}\,u'(0^-).
 \end{equation}
 Equations \eqref{eq:Matching2} and \eqref{eq:Matching11} can be rewritten as

		\begin{equation}\label{eqI:Matching1}
		\left(
		\begin{array}{c}
			u(0^{+}) \\[0.5ex]
			\displaystyle u'(0^{+}) \\
		\end{array}
		\right)= \left(
		\begin{array}{cc}
			\dfrac{1+\lambda_1}{1-\lambda_1}  & 0 \\[0.5ex]
			\dfrac{\lambda_0}{1-\lambda_1^2} & \dfrac{1-\lambda_1}{1+\lambda_1} \\
		\end{array}
		\right)\left(
		\begin{array}{c}
			u(0^{-}) \\[0.5ex]
			\displaystyle {u'}(0^{-}) \\
		\end{array}
		\right),
	\end{equation}
which can be used as a definition of the potential $V_{\ddp}$ for $H_{\ell=0}$ by just replacing $0^\pm$ by $r_0^\pm$. As we discussed above, spherical symmetry forces us to maintain the definition of $H_\ell$ as the Hamiltonian
\begin{equation}
	H^{(0)}_\ell=-\partial_r^2+ \dfrac{\ell(\ell+1)}{r^2}, \quad r\in(0,r_0)\cup (r_0,\infty)
\end{equation} 
equipped with the matching condition
	\begin{equation}\label{eqI:Matching1}
	\left(
	\begin{array}{c}
		u_\ell(r_0^{+}) \\[0.5ex]
		\displaystyle u_\ell'(r_0^{+}) \\
	\end{array}
	\right)= \left(
	\begin{array}{cc}
		\dfrac{1+\lambda_1}{1-\lambda_1}  & 0 \\[0.5ex]
		\dfrac{\lambda_0}{1-\lambda_1^2} & \dfrac{1-\lambda_1}{1+\lambda_1} \\
	\end{array}
	\right)\left(
	\begin{array}{c}
		u_\ell(r_0^{-}) \\[0.5ex]
		\displaystyle {u_\ell'}(r_0^{-}) \\
	\end{array}
	\right),
\end{equation}
for any angular momentum $\ell$, that fixes the space of reduced radial wave functions. When we consider the full radial function $\rho_\ell(r)=u_\ell(r)/r$, we recover the $\tilde\beta$ that shows in Eq.~\eqref{eq:matchG}

 To demonstrate that this matching condition defines self-adjoint operators $H_\ell$, we just need to prove their self-adjointness. We focus on the conditions ensuring that the resulting operator is symmetric. The analysis of the domain of the adjoint operator needed for proving the self-adjointness can be found in Appendix A of \cite{romaniega2020approximation}, where von Neumann’s theory is used. Alternatively, Asorey, Marmo, and Ibort developed a geometrical theory of self-adjoint extensions for Laplace and Dirac operators, and demonstrated that the symmetry condition for these operators is enough to built all the self-adjoint extensions (see Refs. \cite{AIM1,AIMrev}). Their idea can be used to study singular potentials, and was originally developed by Boya and Sudarshan in Ref. \cite{boya}. Let $u_\ell(r),v_\ell(r)\in{L}^2\left((0,r_0)\cup (r_0,\infty)\right)$, and let us denote the scalar product of reduced radial wave functions as
\begin{equation}
	\langle u_\ell,v_\ell\rangle\equiv\int_0^{r_0^-}u_\ell^*(r)v_\ell(r)\,\mathrm{d}r+\int_{r_0^+}^{\infty}u_\ell^*(r)v_\ell(r)\,\mathrm{d}r.
\end{equation}
From the previous definition the obstruction for the operator
$$H_\ell^{(0)}=-\partial_r^2+ \dfrac{\ell(\ell+1)}{r^2},$$
acting on the radial wave functions belonging to a subspace\footnote{As mentioned, this subspace if properly defined in \cite{romaniega2020approximation}.} of ${L}^2\left((0,r_0)\cup (r_0,\infty)\right)$, to be self adjoint is the quantity
\begin{equation}\label{a-14}
\Sigma(u_\ell,v_\ell)\equiv\langle u_\ell,H_\ell^{(0)} v_\ell\rangle-\langle H_\ell^{(0)}u_\ell,v_\ell\rangle.
\end{equation}
It is of note, that the quantity in Eq.~\eqref{a-14} is nothing but the probability current through the spherical shell centered at the origin of radius $r_0$ for a fixed angular momentum. In general, for any pair of functions $u_\ell(r),v_\ell(r)\in{L}^2\left((0,r_0)\cup (r_0,\infty)\right)$ where no matching condition at $r=r_0$ is imposed, the probability flux $\Sigma(u_\ell,v_\ell)$ is non-zero and hence $H_\ell^{(0)}$ is not a self adjoint operator over the appropriate {sub}space  of ${L}^2\left((0,r_0)\cup (r_0,\infty)\right)$. This problem can be solved by restricting the space of reduced radial wave functions through matching or boundary conditions that ensure the cancellation of the boundary probability flux $\Sigma$. Taking into account that the term $\ell(\ell+1)r^{-2}$ obviously verifies
\begin{equation}
	\langle u_\ell,\ell(\ell+1)r^{-2} v_\ell\rangle-\langle \ell(\ell+1)r^{-2}u_\ell,v_\ell\rangle=0,
\end{equation}
then
\begin{equation*}
\Sigma(u_\ell,v_\ell)=	-\left(\langle u_\ell,\partial_r^2 v_\ell\rangle-\langle \partial_r^2 u_\ell,v_\ell\rangle\right).
\end{equation*}
Integration by parts twice enables us two write the probability current $\Sigma(u_\ell,v_\ell)$ as
\begin{widetext}
\begin{equation}\label{a-16}
	\Sigma(u_\ell,v_\ell)=\Phi^\dagger(u_\ell;r_0^-)\mathbb{J}\Phi(v_\ell;r_0^-)-\Phi^\dagger(u_\ell;r_0^+)\mathbb{J}\Phi(v_\ell;r_0^+); \quad \mathbb{J}=\left(\begin{matrix}
		0 & -1 \\
		1 & 0
	\end{matrix}\right);\,\, \Phi(f;r_0^\pm)\equiv \left(\begin{matrix}
	f(r_0^\pm) \\ f'(r_0^\pm)
\end{matrix}   \right),\,\, f=u_\ell,v_\ell
\end{equation}
\end{widetext}
The self-adjointness of $H_\ell^{(0)}$ is only ensured in the domain of reduced radial functions that verify  $\Sigma(u,v)=0$. Taking Eq. \eqref{a-16} into account, $H_\ell^{(0)}$ is self adjoint for domains of functions satisfying a matching condition of the form
\begin{equation}
	\Phi(f,r_0^+)=M\Phi(f,r_0^-)
\end{equation}
as long as the $2\times 2$ complex matrix $M$ satisfies the condition 
\begin{equation}
	M^\dagger\mathbb{J} M=\mathbb{J}.
\end{equation}
It is a straightforward calculation to check that Kurasov's definition of the $\ddp$ potential through the matching condition defined by the matrix
\begin{equation}
	\left(
	\begin{array}{cc}
		\dfrac{1+\lambda_1}{1-\lambda_1}  & 0 \\[0.5ex]
		\dfrac{\lambda_0}{1-\lambda_1^2} & \dfrac{1-\lambda_1}{1+\lambda_1} \\
	\end{array}
	\right),
\end{equation}
leaves invariant the symplectic quadratic form $\mathbb{J}$ given above. Therefore, we conclude that the one-dimensional matching conditions used to define the $\ddp$ potential are valid to define the $\ddp$ spherical shell matching conditions for the reduced radial wave functions. From the matching conditions over the reduced radial functions is then straightforward to implement the matching conditions given in Eq. \eqref{eq:matchG}.

The matching conditions obtained here coincide with the ones found in \cite{kurasov1996distribution}. In this reference the procedure is based on constructing the appropriate theory of distributions for discontinuous functions. Within this context, the matching conditions \eqref{eqI:Matching1} arise when imposing that the Hamiltonian with the $\ddp$ interaction acting on a square integrable function results in another square integrable function. This is explained in the first part of the proof of Theorem 1. It is then proved that for discontinuous functions  the resulting Hamiltonian is also self-adjoint. Although the derivation based on integrating the differential equation presented here leads to the same matching conditions, some technical points described in \cite{kurasov1996distribution} should be considered for the $\delta'$ interaction in order to have a consistent derivation \cite{coutinho1997generalized}. Indeed, in \cite{griffiths1993boundary} the matching conditions for the $n$th derivative of the delta function are also presented integrating the Schr\"odinger equation. However, there are some flaws arising from the discontinuity of the test functions and the final result leads to a operator which is not self-adjoint for odd $n>1$ \cite{coutinho1997generalized,coutinho2012one}.

\subsection{Calculation of the phase shift and Jost function for the $\ddp$ spherical shell.}

As we have indicated, the Jost function of the scattering problem completely determines the self-energy in this context of vacuum fluctuations around classical configurations \eqref{eq:E0}. This function can be easily obtained using the matching conditions \eqref{eq:matchG}. First, note that the radial part of the field can be written as
\begin{equation}\label{eq:sol_scatt}
	\rho_\ell(r)=\left\{ 
	\begin{array}{cc}
		A_1\,j_{\ell}(\omega r)+B_1\,y_{\ell}(\omega r)& \quad r<r_0, \\[1ex]
		A_2\,j_{\ell}(\omega r) +B_2\,y_{\ell}(\omega r)  & \quad r>r_0, \\
	\end{array}
	\right. 
\end{equation}
being $j_{\ell}(x)$ and $y_{\ell}(x)$ the  spherical Bessel functions of the first and second kind respectively
The regularity condition at the origin imposes $B_1=0$. As it is proved in \cite{munoz2019hyperspherical}, the tangent of the phase shift is given by 
$$
\tan{\delta_\ell (\omega) } = -{B_2}/{A_2},
$$
which determines the Jost function \eqref{eq:fJostPhase}. In addition, the solution for $r\to\infty$ defines the
Jost function
\begin{equation}
\rho_\ell(r)= f_\ell(\omega) h_\ell^{(2)}(\omega r)+f^*_\ell(\omega) h_\ell^{(1)}(\omega r),
\end{equation}
being $ h_\ell^{(1,2)} $ the spherical Hankel functions of the first and second kind, respectively. Noting that the Hankel functions are related to the Bessel functions by
\begin{equation*}
h_{\ell }^{(1)}(x)=j_{\ell }(x)+iy_{\ell }(x),\ h_{\ell }^{(2)}(x)=j_{\ell }(x)-iy_{\ell }(x)
\end{equation*}
we conclude that
\begin{equation*}
f_\ell(\omega) = \frac{A_2}{2}+\frac{1}{2} i B_2.
\end{equation*}
The coefficients $\{A_2, B_2\}$  are written imposing in \eqref{eq:sol_scatt} the matching conditions \eqref{eq:matchG} \cite{munoz2019hyperspherical}:
\begin{eqnarray}\label{eq:ExteriorCoefficients}
	\left(\!\!
	\begin{array}{c}
		A_2 \\
		B_2 \\
	\end{array}
	\!\right)\!\!=\!\! \left(
	\begin{array}{cc}
		j_{\ell}(x_0) & y_{\ell}(x_0) \\ [0.5ex]
		j'_{\ell}(x_0) & y'_{\ell}(x_0) \nonumber  \\[0.5ex]
	\end{array}
	\right)^{-1}A_1\,\,\left(
	\begin{array}{cc}
		\alpha & 0 \\ [0.5ex]
		{\tilde \beta}  & \alpha^{-1} \nonumber  \\[0.5ex]
	\end{array}
	\right)\!\! \left(\!
	\begin{array}{c}
		j_{\ell}(x_0) \\ 
		j'_{\ell}(x_0) \\
	\end{array}
	\!\right)\!
\end{eqnarray} 
being $x_0\equiv \omega r_0$ and the derivative defined over the argument.
Note that we can drop any global factor independent of $\omega$ in $f_\ell(\omega)$ since it does not contribute to the energy \eqref{eq:E0}. Finally, the Jost function given in \eqref{jost} is obtained using the relation between the Bessel and Hankel functions with complex arguments $\omega = i\kappa$ and the modified Bessel functions \cite{olver2010nist}.
\section{Consistency checks}\label{appendix:B}
\begin{figure}[h!]
	 \begin{subfigure}
	 {1\columnwidth}
		\includegraphics[width=0.8\textwidth]{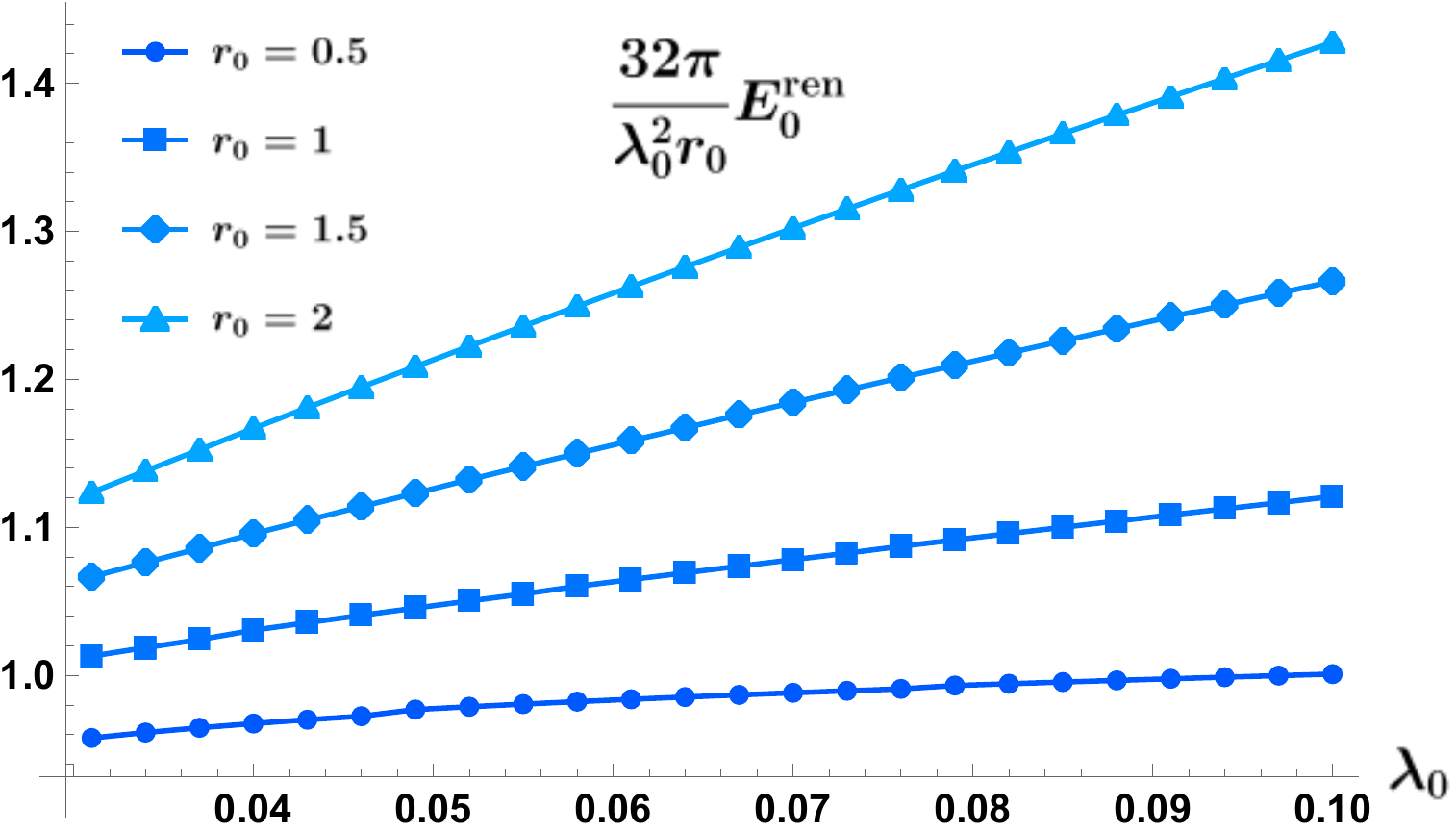}
		\caption{$E_0^\text{ren}$ multiplied by the inverse of $E_0^{(2)}$. Values close to one are expected when the approximation is fair enough.}
		\label{fig:y equals x}
	\end{subfigure}
 \hfill
	\begin{subfigure}
	{1\columnwidth}
		\includegraphics[width=0.8\textwidth]{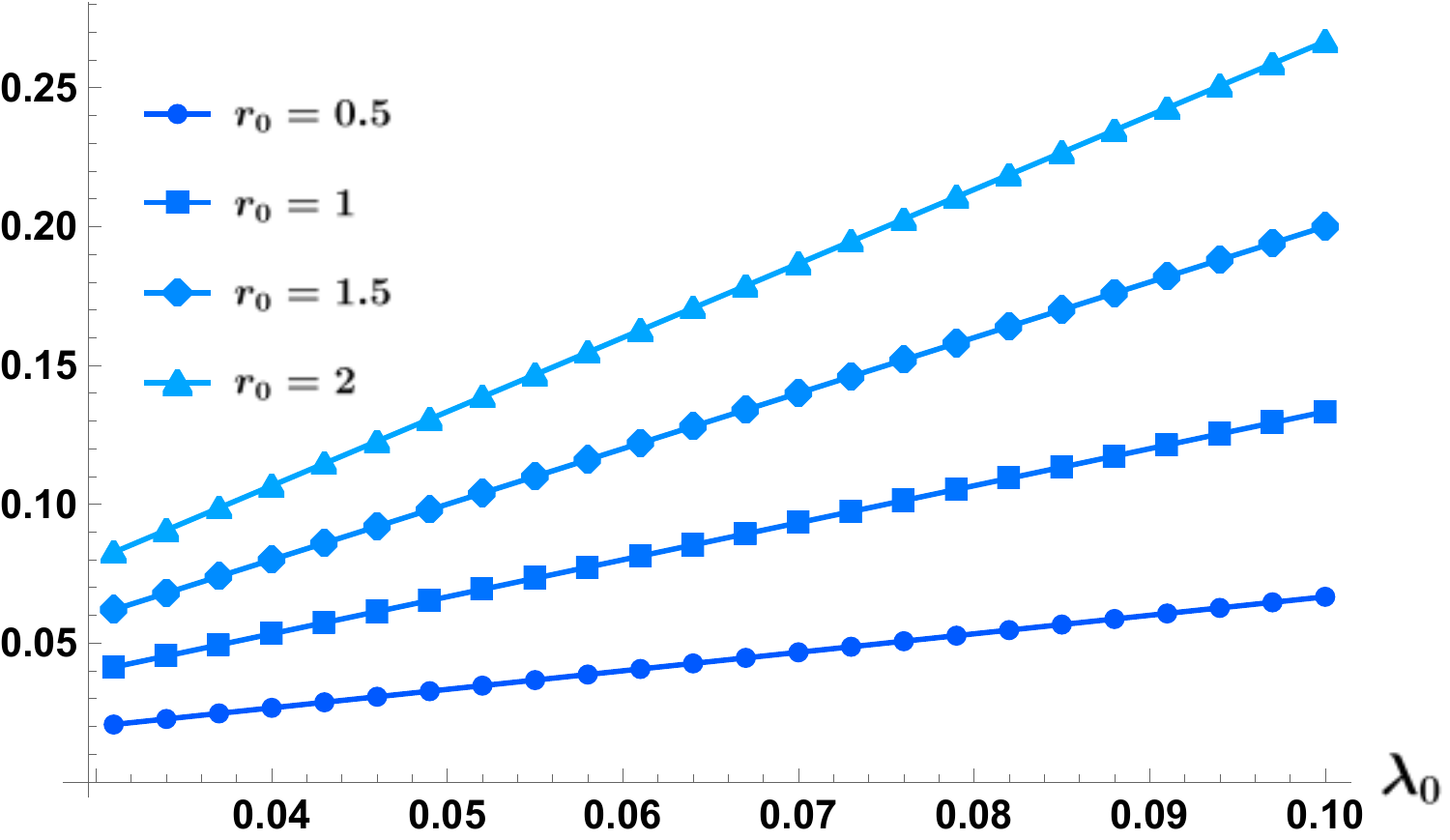}
		\caption{The ratio $E_0^{(3)}/(E_0^{(2)}\zeta(1))$ is evaluated in order to check the relevance of each term.}
		\label{fig:y equals x}
	\end{subfigure}
		\centering
	\caption{Weak limit for the $\delta$-potential, $\lambda_{1}=0$ and small $\lambda_0$. Note that both $E_0^{(2)}$ and $E_0^{(3)}$ are positive quantities.}\label{fig:check_weak_limit}
\end{figure}

In this section we compare our results with previous work. This can be done for particular values of $\{\lambda_0,\ \lambda_1\}$ where our potential simplifies to well known cases.  First, we assume $\lambda_1=0$ and small values of the $\delta$ coupling, i.e. the weak limit for the $\delta$-potential. In \cite{milton2004casimirdelta}  it is proved that, expanding the log in the total energy, an unambiguously finite energy is obtained in second order of the coupling:
\begin{equation}
	E_0^{(2)}= \frac{ r_0}{32 \pi }\lambda_ 0^2.
\end{equation}
Our results are plotted in Fig.~\ref{fig:check_weak_limit}.
Note that the third order term of the expansion is unambiguously divergent. Indeed, this was first proved in \cite{bordag1999ground} and can be seen from our expressions. Specifically, from \eqref{eq:E0as2} we know that the divergence is proportional to $a_2$. For the $\delta$-potential we have already mentioned that $a_2\neq 0$ except for the trivial case. In fact, $a_2$ is proportional to $\lambda_0^3$ as we can see from \eqref{eq:a2}. In particular, the third order term \cite{milton2004casimirdelta} is, 

\begin{equation}\label{eq:E03}
	E_0^{(3)}= \frac{ r_0^2}{24 \pi }\lambda_ 0^3\zeta(1).
\end{equation}
Although the integral is convergent, the sum over the angular momentum is divergent. This is why the Riemann zeta function  $\zeta(z)$  is evaluated at $z=1$. In Fig.~\ref{fig:check_weak_limit} we can see that the result is in good agreement for small values of $\lambda_0$. When the third order term becomes relevant the difference between $E_0^{(2)}$ and $E_0^\text{ren}$ grows larger. Note that in our case the sum is computed until certain $\ell_\text{max}$. In consequence $\zeta(1)$ is only evaluated up to that $\ell_\text{max}$.

We can also verify the case $\lambda_1\to\pm 1$ and $\lambda_0=0$ making use of known
results for the electromagnetic field. In this case we approach the boundary conditions satisfied by the transverse electric (TE) mode and the transverse magnetic (TM) mode  of the electromagnetic field in the presence of a perfectly conducting spherical shell \cite{leseduarte1996complete}. Indeed, due to the spherical symmetry of the system, the electromagnetic problem reduces
	to two independent scalar problems, one for each polarization. The only difference is that for the electromagnetic field there is no contribution from $\ell=0$ \cite{johnson1988invariant}.
In particular, in \cite{leseduarte1996complete} it is found that for the TE mode adding the scalar $\ell=0$ term, $\text{TE}^0$,  the \textit{renormalized} term of the zeta function inside and outside the sphere, 
$2 e_0^\text{ren}$, is
\begin{equation*}
	2 e_0^\text{ren}(\text{TE}^0_\text{in})\simeq 0.00889,\quad 2 e_0^\text{ren}(\text{TE}^0_\text{out})\simeq -0.00326.
\end{equation*}
For the TM mode plus the $\ell=0$ contribution, $\text{TM}^0$, this term inside and outside is
\begin{equation*}
2 e_0^\text{ren}(\text{TM}^0_\text{in})\simeq 0.02805,	\quad 2 e_0^\text{ren}(\text{TM}^0_\text{out})\simeq -0.07223.
\end{equation*}
Note that the sum of the previous four terms would give Boyer's result for a perfectly conducting sphere \cite{boyer1968quantum} if the $\ell=0$ contribution were removed \cite{leseduarte1996complete}.
For each mode we have a scalar problem satisfying Dirichlet ($\text{TE}^0$) or Robin ($\text{TM}^0$) boundary conditions. For the latter, the Robin boundary conditions are the ones in Eq.~\eqref{eq:RobinDirichlet} for $\lambda_0=0$. Consequently, in a system with Robin inside and Dirichlet outside the renormalized energy should be 
\begin{equation*}
e_0^\text{ren}\simeq \dfrac{ 0.02805-0.00326}{2}\simeq 0.012395.
\end{equation*}
With Dirichlet inside and Robin outside
\begin{equation*}
e_0^\text{ren}\simeq \dfrac{ 0.00889-0.07223}{2}\simeq -0.03167.
\end{equation*}
Bearing in mind Eq.~\eqref{eq:RobinDirichlet}, the previous systems can be reached with our potential setting $\lambda_0=0$ and $\lambda_1\to \mp 1$, respectively.
From our code we obtain
\begin{eqnarray*}
\lambda_0=0, \ \lambda_1\to -1,\quad &e_0^\text{ren}\simeq & 0.01241,\\[0.5ex]
\lambda_0=0, \ \lambda_1\to +1,\quad &e_0^\text{ren}\simeq& -0.03166.
\end{eqnarray*}
We want to point out that in these cases $a_2\neq 0$ so the renormalized vacuum energy is not properly defined. Nevertheless, this part of the zeta function can be computed in order to check our findings.


\begin{thebibliography}{10}
	
	\bibitem{bordag2009advances}
	M.~Bordag, G.~L. Klimchitskaya, U.~Mohideen, and V.~M. Mostepanenko.
	\newblock {\em Advances in the Casimir effect}
	\newblock (Oxford Univ. Press, New York, 2009).
	
	\bibitem{milton2009zeropoint}
	K.~A. Milton.
	\newblock {\em The Casimir effect: Physical Manifestations of Zero-point
		Energy} 
	\newblock (World Scientific, Singapore, 2001).
	
	\bibitem{kenneth2006opposites}
	O.~Kenneth and I.~Klich.
	\newblock {\em Phys. Rev. Lett.} \textbf{97}, 160401 (2006).
	
	\bibitem{kenneth2008casimir}
	O.~Kenneth and I.~Klich.
	\newblock {\em Phys. Rev. B} \textbf{78}, 014103 (2008).
	
	\bibitem{rahi2009scattering}
	S.~J. Rahi, T.~Emig, N.~Graham, R.~L. Jaffe, and M.~Kardar.
	\newblock {\em Phys. Rev. D} \textbf{80}, 085021 (2009).
	
	\bibitem{PhysRevLett.82.3948}
	I.~Brevik, V.~N. Marachevsky, and K.~A. Milton.
	\newblock {\em Phys. Rev. Lett.} \textbf{82}, 3948--3951 (1999).
	
	\bibitem{Barton_1999}
	G.~Barton.
	\newblock {\em J. Phys. A Math. Gen.} \textbf{32}, 525--535
	(1999).
	
	\bibitem{ROMEO2005309}
	A.~ Romeo and K.~A. Milton.
	\newblock {\em Phys. Lett. B} \textbf{621}, 309--317 (2005).
	
\bibitem{marachevsky2001casimir}
	V. N. Marachevsky.
	\newblock {\em Phys. Scr.} \textbf{64}, 205 (2001).
	
	\bibitem{boyer1968quantum}
	T.~H.~Boyer.
	\newblock {\em Phys. Rev.} \textbf{174}, 1764 (1968).
	
	\bibitem{deraad1981casimir}
	L.~L.~DeRaad~Jr and K.~A.~Milton.
	\newblock {\em Ann. Phys.} \textbf{136}, 229--242 (1981).
	
	\bibitem{PhysRevE.55.4207}
	K.~A.~Milton and Y.~Jack Ng.
	\newblock {\em Phys. Rev. E} \textbf{55}, 4207--4216 (1997).
	
	\bibitem{cavero2005casimir}
	I.~Cavero-Pelaez and K.~A.~Milton.
	\newblock {\em Ann. Phys.} \textbf{320}, 108--134 (2005).
	
	\bibitem{bordag1999ground}
	M.~Bordag, K.~Kirsten, and D.~Vassilevich.
	\newblock {\em Phys. Rev. D} \textbf{59}, 085011 (1999).
	
	\bibitem{kirsten2001spectral}
	K.~Kirsten.
	\newblock {\em Spectral Functions in Mathematics and Physics}.
	\newblock (Chapman \& Hall/CRC, Boca Raton, 2001).
	
	\bibitem{vassilevich2003heat}
	D.~Vassilevich.
	\newblock Heat kernel expansion: user's manual.
	\newblock {\em Phys. Rept.} \textbf{388}, 279--360 (2003).
	
	\bibitem{cavero2021casimir}
	I.~Cavero-Pel{\'a}ez, J.M. Munoz-Castaneda, and C.~Romaniega.
	\newblock {\em Phys. Rev. D} \textbf{103}, 045005 (2021).
	
	\bibitem{romaniega2021repulsive}
	C.~Romaniega.
	\newblock {\em Eur. Phys. J. Plus} \textbf{136}, 327 (2021).
	
	\bibitem{milton2008local}
	K.~A.~Milton, I~Cavero-Pelaez, and K~Kirsten.
	\newblock In {\em The Eleventh Marcel Grossmann Meeting: On Recent Developments
		in Theoretical and Experimental General Relativity, Gravitation and
		Relativistic Field Theories (In 3 Volumes)}, pages 2727--2745. World
	Scientific, 2008.
	
	\bibitem{klich1999casimir}
	I.~Klich.
	\newblock {\em Phys. Rev. D} \textbf{61}, 025004 (1999).
	
	\bibitem{milton1999mode}
	K.~A.~Milton, A.~V.~Nesterenko, and V.~V.~Nesterenko.
	\newblock {\em Phys. Rev. D} \textbf{59}, 105009 (1999).
	
	\bibitem{brevik1982electrostriction}
	I. Brevik.
	\newblock {\em J. Phys. A} \textbf{15}, L369 (1982).
	
\bibitem{brevik1982casimir}
	I. Brevik and H. Kolbenstvedt.
	\newblock {\em Ann. Phys.} \textbf{143}, 179 (1982).
	
	\bibitem{brevik1985attractive}
	I. Brevik and H. Kolbenstvedt.
	\newblock {\em Can. J. Phys.} \textbf{63}, 1409 (1985).
	
\bibitem{brevik1988casimir}
	I. Brevik and G. Einevoll.
	\newblock {\em Phys. Rev. D} \textbf{37}, 2977 (1988).
		
	\bibitem{milton2004casimirdelta}
	K.~A.~ Milton.
	\newblock {\em J. Phys A Math. Gen.} \textbf{37}, 6391 (2004).
	
	\bibitem{weyl1912asymptotische}
	Hermann Weyl.
	\newblock {\em Math. Ann.} \textbf{71}, 441--479 (1912).
	
	\bibitem{kurasov1996distribution}
	P.~Kurasov.
	\newblock {\em J. Math. Anal. Appl.} \textbf{201}, 297 (1996).
	
	\bibitem{gadella2009bound}
	M.~Gadella, J.~Negro, and L.~M.~Nieto.
	\newblock {\em Phys. Lett. A} \textbf{373}, 1310--1313 (2009).
	
	\bibitem{munoz2015delta}
	J.~M. Mu{\~n}oz-Casta{\~n}eda and J.~M. Guilarte.
	\newblock {\em Phys. Rev. D} \textbf{91}, 025028 (2015).
	
	\bibitem{albeverio2000singular}
	S.~Albeverio and P.~Kurasov.
	\newblock {\em Singular perturbations of differential operators: solvable
		Schr{\"o}dinger-type operators}
	\newblock (Cambridge Univ. Press, Cambridge, 2000).
	
	\bibitem{martin2022solvable}
	A.~Mart{\'\i}n-Mozo, L.~M.~Nieto, and C.~Romaniega.
	\newblock {\em Eur. Phys. J. Plus} \textbf{137}, 1--24 (2022).
	
	\bibitem{munoz2019hyperspherical}
	J.~M. Mu{\~n}oz-Casta{\~n}eda, L.~M. Nieto, and C.~Romaniega.
	\newblock {\em Ann. Phys.} \textbf{400}, 246 (2019).
	
	\bibitem{bordag1996vacuum}
	M.~Bordag and K.~Kirsten.
	\newblock {\em Phys. Rev. D} \textbf{53}, 5753 (1996).
	
	\bibitem{taylor2006scattering}
	J.~R. Taylor.
	\newblock {\em Scattering theory: the quantum theory of nonrelativistic
		collisions}.
	\newblock (Dover Publications, New York, 2006).
	
	\bibitem{olver2010nist}
	F.~W.~J. Olver, D.~W. Lozier, R.~F. Boisvert, and C.~W. Clark.
	\newblock {\em NIST Handbook of Mathematical Functions}
	\newblock (Cambridge Univ. Press, Cambridge, 2010).
	
	\bibitem{bordag1999heat}
	M.~Bordag and D.~V.~Vassilevich.
	\newblock {\em J. Phys. A Math. Gen.} \textbf{32}, 8247	(1999).
	
	\bibitem{cavero2006local}
	I.~Cavero-Pel{\'a}ez, K.~A.~Milton, and J.~Wagner.
	\newblock {\em Phys. Rev. D} \textbf{73}, 085004 (2006).
	
	\bibitem{barton2004casimir}
	G.~Barton.
	\newblock {\em J. Phys. A Math. Gen.} \textbf{37}, 3725 (2004).
	
	\bibitem{li2019casimir}
	Y.~Li, K.~A.~Milton, X.~Guo, G.~Kennedy, and S.~A. Fulling.
	\newblock {\em Phys. Rev. D} \textbf{99}, 125004 (2019).
	
	\bibitem{romaniega2020approximation}
	C.~Romaniega, M.~Gadella, R.~M.~Id Betan, and L.~M.~Nieto.
	\newblock {\em Eur. Phys. J. Plus} \textbf{135}, 372 (2020).
	
	\bibitem{leseduarte1996complete}
	S.~Leseduarte and A.~Romeo.
	\newblock {\em Ann. Phys.} \textbf{250}, 448--484 (1996).
	
	\bibitem{milton2004casimir}
	K.~A. Milton.
	\newblock {\em J. Phys. A Math. Gen.} \textbf{37}, R209 (2004).

\bibitem{santagata}
	A. Santagata
	\newblock{{\em Coulomb Phases: from Graphene to Quark Confinement.}}
	\newblock{(PhD dissertation, Zaragoza University 2014).}
	
\bibitem{griffiths1993boundary}
	 D. J. Griffiths.
	\newblock {\em J. Phys. A Math. Gen.} \textbf{26}, 2265 (1993).

\bibitem{cavero2008NonConntG}
	I.~Cavero-Pel{\'a}ez, K.~A.~Milton, P.~Parashar, and K.~V.~Shajesh.
	\newblock {\em Phys. Rev. D} \textbf{78}, 065018 (2008).

\bibitem{AIM1}
M.~Asorey, A.~Ibort and G.~Marmo,
	\newblock {\em Int. J. Mod. Phys. A} \textbf{20}, 1001-1026 (2005).

\bibitem{AIMrev}
M.~Asorey, A.~Ibort and G.~Marmo,
\newblock {\em Int. J. Geom. Meth. Mod. Phys.} \textbf{12},6, 156100 (2015).

\bibitem{boya}
L.~J.~Boya and E.~C.~G.~Sudarshan,
\newblock {\em Int. J. Theor. Phys.} \textbf{35}, 1063-1068 (1996).

\bibitem{coutinho1997generalized}
	F. A. B. Coutinho, Y. Nogami, and J. F. Perez.
	\newblock {\em J. Phys. A Math. Gen.} \textbf{30}, 3937 (1997).

\bibitem{coutinho2012one}
	F. A. B. Coutinho, Y. Nogami, and F. M. Toyama.
	\newblock {\em Can. J. Phys.} \textbf{90}, 383 (2012).
	
\bibitem{johnson1988invariant}
	B. R. Johnson.
	\newblock {\em Appl. Opt.} \textbf{27}, 4861 (1988).

\end{thebibliography}
\end{document}